\documentclass [dvips, 12pt] {article} 
\usepackage {graphicx}
\usepackage {pifont}
\newcommand {\pion} {\ensuremath {{\pi^0}} }

\newcommand*{\Approx}[1]{\ensuremath {\approx #1}}

\newcommand {\cd} {\normalsize \textcircled {\em \scriptsize d}}
\newcommand {\cu} {\normalsize \textcircled {\em \footnotesize u}}
\newcommand {\cs} {\normalsize \textcircled {\em \footnotesize s}}
\newcommand {\cc} {\normalsize \textcircled {\em \footnotesize c}}
\newcommand {\cb} {\normalsize \textcircled {\em \scriptsize b}}

\newcommand {\cdb} {\normalsize \textcircled {\em \scriptsize   $\bar{d}$}}
\newcommand {\cub} {\normalsize \textcircled {\em \footnotesize $\bar{u}$}}
\newcommand {\csb} {\normalsize \textcircled {\em \footnotesize $\bar{s}$}}
\newcommand {\ccb} {\normalsize \textcircled {\em \footnotesize $\bar{c}$}}
\newcommand {\cbb} {\normalsize \textcircled {\em \scriptsize $\bar{b}$}}

\begin {document}

\begin {center}
{\bf \large The exact quantization of the CLEO and BELLE data \\ 
for $D$ mass differences by the harmonic quarks and oscillators}
     \par\bigskip
 
{\Large Oleg~A.~Teplov}
     \par\smallskip 
Institute of metallurgy and materiology of the Russian Academy of Science, 
Moscow.
\par e-mail: teplov@ultra.imet.ac.ru 
\end {center}
\par\bigskip
\begin{abstract}
The harmonic quarks and their complete oscillators are successfully used
for the exact quantitative results and an explanation of 
the $D$ meson mass differences. 
For the first time it is shown that the mass differences of the charmed mesons
with the same flavors are strictly quantized by the rest masses 
of harmonic quarks in both "free" state and complete oscillator state.
The  $D^*(2007)^0$ $\rightarrow$ $D^0$ transition  has 
the following neutral quark group: $d$-oscillator + $u^0$-quark with rest energy 142.124 MeV.
It is argued that it is the unique model solution 
within the 4$\sigma$ experimental interval.
The $D_s^{*+}$ - $D_s^+$ mass difference is quantized 
by the simplest quark reaction with the difference energy 143.756 MeV.
The energy of the $D_{sJ}$(2460)$^+$ $\rightarrow$ $D_s^+$ transition 
explains by the decay of the one complete 
$s$-oscillator with the energy 491.33 MeV.
The full agreement with experimental data of CLEO and BELLE is obtained.  
The problems of a quark shells and the quantization 
by quark rest masses are discussed.   

\end{abstract}

\section {Introduction}

In this article the harmonic quarks and their complete 
oscillators are used for a decoding of the mass differences 
of the $D$ mesons. 

The articles~\cite{my1,my2,my3} have shown that the harmonic quarks 
are powerful and effective tool for study of hadronic structures. 
This tool works equally well with both ground energy levels, and resonances.
The harmonic quarks and oscillators allow decrypt the mass spectrum of charmed mesons~\cite{my1} completely and with great precision.
Simultaneously, there was established that the model harmonic spectrum 
is not sensitive to the isospin.
In other words, the model harmonic spectrum is a degenerate spectrum 
in relation to this property. 
At the same time, the levels of harmonic spectrum 
of charmed mesons are only quantized by masses of the harmonic quarks 
and their oscillators of accessible flavors. All levels are the sums 
of the rest masses of the harmonic quarks and oscillators. 
Therefore we may suppose that both real masses of charmed hadrons
and the mass differences between them can also be quantized by similar 
combinations.
The investigation of this assumption can help us resolve the open questions
of the hadronic physics, such as the hadronization stage, 
or the electromagnetic split of a levels.

The present paper can be seen as the continuation of the previous 
work~\cite{my1}, with the only difference, that it will be oriented 
to a study of the energy structure of the real transitions 
in charmed mesons.
With the help of the most precise experimental 
data~\cite{cleo, belle, cleo145, cleo143} we shall discover the groups 
of the harmonic quarks 
and oscillators which will be in full agreement with experience data. 

Here we shall use the same labels as in~\cite{my1}  
with the standard notation for the harmonic quarks ($d$, $u$, $s$, $c$), 
and circle-enclosed quarks of the complete harmonic oscillators (\cd, \cu, \cs).    

The masses of harmonic quarks and complete harmonic 
oscillators are given in the table 1.

\begin{center}
Table 1. The masses of harmonic quarks and their complete oscillators.  

  \medskip\small 
  \begin{tabular}{|c|c|c|c|}
    \hline
Quark & Quark mass, & Harmonic & Oscillator energy, \\
       &  MeV & oscillator  & MeV  \\
    \hline
   $d$ & 28.8106 & \cd \cdb  & 36.683    \\
\hline
   $u$ & 105.441 & \cu \cub & 134.251  \\
\hline
   $s$ & 385.891 & \cs \csb & 491.332  \\
\hline
   $c$ & 1412.28  & \cc \ccb & 1798.17 \\
\hline
   $b$ & 5168.7   & \cb \cbb & 6581.0    \\
\hline

  \end{tabular}

\end{center}

\par \bigskip

\section {Transitions between $D$ mesons}

\quad The most precise experimental data about mass differences 
of the charmed mesons are given in the table 2.
In general we shall use the consistent data of PDG~\cite{parti}, 
and also the most precise data of CLEO collaboration 
for transitions $D_s^{*+}$ to $D_s^+$~\cite{cleo143} and, 
especially, $D^{*0}$ to $D^0$ and $D^{*+}$ to $D^0$~\cite{cleo, cleo145}. 
The data for 5 transitions in table 2 are given in decreasing order of precision.
All transitions occur without change of the strangeness 
with formation of stable $D$ mesons at decays of the resonances.  

Decay $D_s^+$ to $D$ will not be considered, since it arises from 
a weak interaction.
\pagebreak
{      
\begin{center}

Table 2. The experimental data for transitions between $D$ mesons
 with the same flavors. 
  \medskip\small 
  \begin{tabular}{|c|c|c|c|}
    \hline
 Transition & Energy of & Main channels & Probability \\
       & transition, MeV & of decay & of decay, \% \\
    \hline
 $(1^-)$$D^*(2010)^+$$\rightarrow$$(0^-)$$D^0$ &$145.412\pm0.002
\pm0.012$~\cite{cleo145}& $D^{*+}$ $\rightarrow$ $D^0\pi^+$ & 68 \\
 & 145.421 $\pm$ 0.010~\cite{parti} &  & \\
\hline
 $(1^-)$$D^*(2007)^0$$\rightarrow$$(0^-)$$D^0$ &$142.12\pm0.05
\pm0.05$~\cite{cleo}& $D^{*0}$ $\rightarrow$ $D^0\pi^0$ & 62 \\
  &  142.12 $\pm$ 0.07~\cite{parti} & $D^{*0}$ $\rightarrow$ $D^0\gamma$ & 38 \\  
\hline
 $(1^-)$$D^*(2010)^+$$\rightarrow$$(0^-)$$D^+$ & $140.64 \pm 0.08 
\pm 0.06$~\cite{cleo}& $D^{*+}$ $\rightarrow$ $D^+\pi^0$ & 31 \\
 & 140.64 $\pm$ 0.10~\cite{parti} & $D^{*+}$ $\rightarrow$ $D^+\gamma$ & 1.6 \\
\hline
 $(1^-)$$D_s^*(2112)^+$$\rightarrow$$(0^-)$$D_s^+$ & $143.76 \pm 0.39
 \pm 0.4$~\cite{cleo143}& $D_s^{*+}$ $\rightarrow$ $D_s^+\gamma$ & 94 \\
 & 143.8 $\pm$ 0.4~\cite{parti}&$D_s^{*+}$ $\rightarrow$ $D_s^+\pi^0$  & 6 \\
\hline

$(1^+)$$D_{sJ}$(2460)$^+$$\rightarrow$$(0^-)$$D_s^+$  & 491.4 $\pm$ 0.9 $\pm$ 1.5~\cite{belle} & $D_{sJ}^+$$\rightarrow$$D_s^+\gamma$  & seen  \\
 &  491.0 $\pm$ 1.2~\cite{parti}(FIT)& $D_{sJ}^+$$\rightarrow$$D_s^+\pi^+\pi^-$ &  \\
 & 491.3 $\pm$ 1.4~\cite{parti}(AVR)& & \\
\hline

  \end{tabular}

\end{center}
}

\par

The most precise data was obtained by collaboration CLEO~\cite{cleo, cleo145, cleo143}. 
We can calculate the energies of the same transitions (see table 3),
using the model harmonic spectrum for the charmed mesons~\cite{my1}
with the aforementioned model restriction of not considering electromagnetic
split.

{
      
\begin{center}
Table 3. The energy differences of harmonic levels and their quark representation. 
  \medskip\small 
  \begin{tabular}{|c|c|c|c|}
    \hline
 Transition &  Difference between& Quark representation\\
       &  harmonic levels, MeV~\cite{my1} & of difference \\
    \hline
$(1^-)$$D^*$(2010)$^+$$\rightarrow$$(0^-)$$D^0$ & 143.756 $\pm$ 0.007
 & $u\bar{u}$ $\rightarrow$ $\cu^0$ \\
\hline
 $(1^-)$$D^*(2007)^0$$\rightarrow$$(0^-)$$D^0$ & 143.756 $\pm$ 0.007 & $u\bar{u}$ $\rightarrow$ $\cu^0$ \\
    
\hline
 $(1^-)$$D^*(2010)^+$$\rightarrow$$(0^-)$$D^+$ &  143.756 $\pm$ 0.007 & $u\bar{u}$ $\rightarrow$ $\cu^0$ \\
\hline
 $(1^-)$$D_s^*(2112)^+$$\rightarrow$$(0^-)$$D_s^+$ & 141.86 $\pm$ 0.007 & 1/2(\cs\csb+\cu\cub)$\rightarrow$ \cu\cub + \cd\cdb \\

\hline
$(1^+)$$D_{sJ}$(2460)$^+$$\rightarrow$$(0^-)$$D_s^+$  & 491.33 $\pm$ 0.025 & \cs\csb $\rightarrow$ 0 \\
\hline

  \end{tabular}

\end{center}

}
\par

The data in table 3 is clearly showing to us the limits 
of the model spectrum with respect to isospin.
Only transitions between $D_s$ singlet states are simple 
and have single meanings.
There are two important coincidences that should be noted
when comparing the experimental and model data in tables 2 and 3.

{\em First of all}, the experimental and model 
data for transition $D_{sJ}$(2460)$^+$ $\rightarrow$ $D_s^+$ 
agree completely:
 491.3 $\pm$ 1.4 (PDG, AVERAGE)\, and \,491.33 $\pm$ 0.025 MeV.

This can only mean one thing:
there's an annihilation of a neutral harmonic oscillator \cs\csb \,
in the decay $D_{sJ}$(2460)$^+$ to $D_s^+$. 
Apparently, $D_{sJ}$(2460)$^+$ just loses one neutral shell 
in its structure.
With the annihilation of a neutral strange oscillator 
491.33 MeV is released and it is subsequently used to form 
the neutral $\gamma$ (or a group $\pi^+\pi^-$) and the some part 
goes to kinetic energy.

{\em Secondly}, the energy of the model quark reaction $u\bar{u}$ $\rightarrow$  
$\cu^0$ completely agrees with the experimental energy of the transition 
$D_s^*$(2112)$^+$ $\rightarrow$ $D_s^+$:

143.756 $\pm$ 0.007 and 143.76 $\pm$ 0.39 $\pm$ 0.4 MeV (CLEO) 
respectively.
Hence, the quark reaction $u\bar{u}$ $\rightarrow$  $\cu^0$ 
really occurs at decay $D_s^*$(2112)$^+$, instead of the model 
reaction with almost equal energy which is given in table 3.  
Quark reaction of this kind was considered earlier in~\cite{my1}.

These two facts support our guess about a quantization of 
the $D$ meson transitions by the quark masses. 

These transitions have a lot in common:
they both happen between singlet states ($D_s$), 
the spin decreases on 1 in both of them and they both generate $D_s^+$.
Therefore it can be said that the transitions $D_s^+$ into excited states 
(inverse transitions are experimentally observed)
are strictly quantized by harmonic quark masses.

Let's get to the next problem.
Whether the most precise experimental data in table 2
can be similarly represented by the rest masses of quark groups?

All unconsidered mass differences relate to the transitions between 
the $D$ and the $D^*$ doublet states.
The three of them are experimentally observed, and
the fourth transition $D^{*0}$ $\rightarrow$ $D^+$ can only be caused 
by weak coupling and is not considered here. 
Only one of them -- $D^*$(2007)$^0$ - $D^0$ (see tab.2) 
with energy 142.12 MeV is similar to the transitions examined above.  
The decay of the $D^*$(2007)$^0$, moreover, with a high probability 
(38 \%) has the mode $D^0\gamma$\,, while the decay $D^{*+}$ 
has the mode $D^+$$\gamma$ with a much smaller probability (1.6\%).  

The energy of transition $D^*(2007)^0$ $\rightarrow$ $D^0$ is quantizing 
excellently. 
One just has to subtract from it the energy of the oscillator \cd\cdb \, 
from it:
 142.12 - 36.68 = 105.44 (MeV) to obtain a rest mass of 
$u$ quark with 5 decimal digits precision! Thus the mass difference 
$\Delta$$M$ $\equiv$ $M_{D^*(2007)^0}$ - $M_{D^0}$ can be written as:
\begin{equation}\label{M142}
   {\Delta}M = M_u + M_{\cd\cdb} = 142.124 \pm 0.007 (MeV) 		
\end {equation}
\par

There is only two neutral quark groups in the energy range 
from the oscillator \cu\cub \, to 145 MeV :
 $\cu^0$ + 2\cd\cdb \,
 and  $u^0$ + \cd\cdb \,,
with respective energies of 140.49 and 142.12 MeV.
(The first value agrees with mass difference $D^*(2010)^+$ - $D^+$:
 140.64 $\pm$ 0.10~\cite{parti}.) 
Thus, the quark representation of the $D^*(2007)^0$ - $D^0$ mass difference 
can only be $u^0$ + \cd\cdb, if we are considering simple neutral quark groups.
However, it's possible to imagine rather complicated quark groups 
to account for the difference and therefore all combinations 
were computed up to the maximum energy (488.98 MeV).
This energy is the mass of ${D^*(2007)^0}$ without the masses of valence 
quarks ($c\bar{u}$), and, above this energy, difference quark groups 
can be formed only with additional energy (vacuum fluctuations), 
which we shall not consider.
The calculations were done for 4$\sigma$ probability interval, 
which is equal to $\pm$0.28 MeV~\cite{parti}.
It is necessary to exclude the groups with one $s$ quark from analysis.
The groups found are given in table 4.  

\begin{center}
Table 4. The difference quark groups within the 142.12 $\pm$ 
0.28 MeV interval for $D^*(2007)^0$.  

  \medskip\small 
  \begin{tabular}{|c|c|c|}
    \hline
Quark group & Energy group, & Deviation  \\
 &  MeV & from 142.12, MeV     \\
    \hline
   $u^0$\cd\cdb & 142.124 & 0.0036    \\
\hline
  7\cd\cdb \, + 7$d$ - 3$u$ & 142.131 & $ 0.0113$   \\
\hline

  \end{tabular}

\end{center}

It's possible to construct only one more difference quark group
within the 4$\sigma$ interval for the $D^*(2007)^0$ - $D^0$ 
experimental error ($\sigma$ = 0.07 MeV~\cite{parti}) -- but 
there are many reasons not to consider it seriously. 
Here are some of them:  color suppression, charges of reaction 
and phase volume suppression.    
For example, there can not be 7 harmonic coupled quarks (\cd \, or \cdb) 
without two of them having the same color and spin. 
Therefore, we have {\bf \em the unique solution} in the 4$\sigma$ interval.
The existence of a neutral mass combination 
$u^0$ was assumed earlier in~\cite{my1}.
The same approach can be used for study of other mass differences. 

Precise equality of the experimental 
and model quantities is the real success 
of the harmonic quark  model. 

There are two more mesons in the $D$ meson spectrum 
with similar mass difference:
$D_{sJ}^*$(2317)$^+$ and $D_{sJ}$(2460)$^+$, which have 
harmonic levels with energies 2318 and 2460 MeV~\cite{my1}. 
Quark representations for these levels are $c$ + $s\bar{s}$ + \cu\cub \, 
and \cc\ccb \,+ \cs\csb \,+ \cu\cub \,+ \cd\cdb. 
The energy difference of these levels are also equal to \cd\cdb \,+ $u^0$.
This result can be easily achieved using first proposition ~\cite{my4}.
The experimental mass difference $D_{sJ}$(2460)$^+$ and 
$D_{sJ}^*$(2317)$^+$ is equal to 141.9 $\pm$ 1.6 ~\cite{parti},
which is also agrees well with the energy of the model quark group. 
It should be noted that $D_{sJ}$(2460)$^+$ contains the decay mode
 $D_{sJ}^*$(2317)$^+$ with photon. It is necessary to mark, 
that this annihilation group is not the single solution because 
of large experimental uncertainty. 

Thus, these 4 transitions can be simply interpreted by means of 
the rest masses of harmonic quarks and oscillators, 
and this interpretaion is in the full accordance 
with the best experimental data.
All these transitions are accompanied by change of $J$ on 1,
and they have the $\gamma$ mode in the main channels of the decay.

The obtained data are presented in the final table 5. 

\pagebreak

{      
\begin{center}
Table 5. The summed data for the mass differences of the $D$ mesons. 
  \medskip\small 
  \begin{tabular}{|c|c|c|}
    \hline
 Transition & Quark group & Energy of group \\
       & of transition & or reaction, MeV  \\
    \hline

\hline
 $D^*(2007)^0$$\rightarrow$$D^0$ & \cd$u^0$\cdb & $142.124 \pm0.007$  \\
  
\hline
 $D_s^*(2112)^+$$\rightarrow$$D_s^+$ & $u\bar{u}$ - $\cu^0$ &
 $143.756 \pm 0.007$  \\

\hline

$D_{sJ}(2460)^+$$\rightarrow$$D_s^+$  & \cs\csb & 
$491.33 \pm 0.024$  \\

\hline
\hline
$D_{sJ}(2460)^+$$\rightarrow$$D_{sJ}(2317)^+$ & \cd$u^0$\cdb & 
$142.124 \pm0.007$  \\
ONE \quad  OF & A POSSIBLE & WAY  \\ 

\hline
 $D^*(2010)^+$$\rightarrow$$D^+$ & \cd\cd$\cu^0$\cdb\cdb &
 $140.49 \pm 0.007$ \\
IT \, REQUIRES & ADDITIONAL & EXAMINATION  \\\hline

  \end{tabular}

\end{center}
}

\par

\section {Discussion of Results}

We established that several transitions in the charmed meson spectrum 
are strictly quantized. That wasn't unexpected, it should be so. 
However, quantization of mass differences by 
the harmonic quark rest masses is a new unexplored phenomenon. 
Both the "free" harmonic quarks and the harmoniously 
bound quarks both participate in this quantization.
What does it mean? Why are there only rest masses of quarks?
It means that there is no active motion inside the group 
of the annihilating quarks.
The quark group manifests itself as a something which is whole 
and annihilates as a whole to the $\gamma$ or $\pion$. 

There are two possible explanations of such a phenomenon:
either the quark group is forming just before annihilation, 
or the group exists as a shell inside the meson.
In the table 4 the second option the \cs\csb \, item may be 
the shell, though it is still an open question.
The transitions \cs\csb \, $\rightarrow$ 0 and 
$u\bar{u}$ $\rightarrow$ $\cu^0$ give us a certain understanding
of quark confiquration of the corresponding mesons.
With the 142.12 MeV group there is no such clarity. 
 
All four transitions between doublet states $D^*$ and $D$ cannot be 
exactly quantized 
by quark masses  since not more than 3 of them can be independent.  
If an electromagnetic split in a doublet state is determined 
by the other circumstances
then only one transition may be exactly quantized 
by quark masses, and 
will define the mutual position of the doublets on a mass scale.  

We will name this quantized transitions as {\bf \em the basic transitions}, 
and other transitions as {\bf \em additional}.

The three of these transitions are experimentally observed, 
an the fourth transition $D^{*0}$ $\rightarrow$ $D^+$ can only be caused 
by a weak interaction and not considered here.
One basic transition is the quark annihilation reaction 
\cd$u^0$\cdb \, $\rightarrow$ 0 agrees precisely 
with CLEO~\cite{cleo} experimental data.
The other basic transitions, if they exists, are not detected.
Although the one quark group of the mass difference 
$D^*(2010)^+$ - $D^+$ 
agree with experimental data 140.64~\cite{parti}.

Transitions for singlet states are unambiguous and therefore 
we  observe 3 simply interpreted transitions between 
the strange charmed mesons.
Two of them agree with a model spectrum~\cite {my1}.
In the third transition we observe the quark reaction 
($u\bar{u}$ $\rightarrow$ $\cu^0$) which is simpler 
than it follows from model spectrum~\cite {my1}.

For three transitions $D_{sJ}(2460)^+$ $\rightarrow$ $D_{sJ}(2317)^+$, 
$D_{sJ}(2460)^+$ $\rightarrow$ $D_s^+$ and $D_{sJ}(2317)^+$  
$\rightarrow$ $D_s^+$ repeat the situation in which are analogy with 
the transitions between the doublet states. Only two transitions can be 
the basic transitions, the third mass difference is by their consequence. 
The author considers the $D_{sJ}(2460)^+$ $\rightarrow$ $D_s^+$ 
is the basic  transition. The second basic transition are perhaps 
the $D_{sJ}(2460)^+$ $\rightarrow$ $D_{sJ}(2317)^+$, 
but it is preliminary conclusion.  

The results of this work can use for the correction of some $D_s$ 
mass data.   
The relative error for quark masses is \Approx 5 times less 
than the experimental error for the masses of  $D$ mesons. 
With the help of $D_s^+$ experimental mass and 
precise energies of the harmonic groups (table 5), 
this makes possible to redefine the masses of several excited $D_s$ states
and, moreover, to decrease their mass errors up to $\pm$0.5 MeV.

The obtained results are given in the table 6.

\begin{center}
Table 6. The calculated masses of strange charmed mesons.  

  \medskip\small 
  \begin{tabular}{|c|c|c|}
    \hline
Meson & Experimental meson mass, & Calculated meson mass,  \\
   &  MeV~\cite{parti} &MeV (this work)  \\
    \hline
   $D_s^+$ & 1968.3 $\pm$ 0.5 &  -   \\
\hline
 $D_s^{*+}$  & 2112.1 $\pm$ 0.7 & 2112.1 $\pm$ 0.5   \\
\hline
   $D_{sJ}^*$(2317)$^+$ & 2317.4 $\pm$ 0.9 & 2317.5 $\pm$ 0.5 \\
\hline
 $D_{sJ}$(2460)$^+$  & 2459.3 $\pm$ 1.3 & 2459.6 $\pm$ 0.5   \\
\hline

  \end{tabular}

\end{center}

The author implyes, that all obtained results can be directly 
carried over to the corresponding antiparticles.
The energy of most precisely measured transition $D^+$ in $D^0$ 
can also be calculated from the model, but this calculation 
and explanation of the results are beyond the scope of this work.

\section {Conclusion}

\quad \, It's discovered that the mass differences
of the $D$ mesons are rigorously quantized by the rest masses
of harmonic quarks. 

The harmonic quarks and their derivatives 
are the main building blocks for the hadrons.

\end {document}